\documentclass[aps,prl,reprint,superscriptaddress,amsmath,amssymb]{revtex4-1}
\pdfoutput=1

\usepackage[utf8]{inputenc}
\usepackage{graphicx,hyperref,siunitx}
\usepackage[all]{hypcap}

\newcommand*{\comm}[2]{\left[ #1,#2 \right]}
\newcommand*{\acomm}[2]{\left\{ #1,#2 \right\}}
\newcommand*{\im}{\, \mathrm{Im}}
\newcommand*{\re}{\, \mathrm{Re}}

\renewcommand*{\eqref}[1]{Eq.~(\ref{#1})}
\newcommand*{\Eqref}[1]{Equation~(\ref{#1})}
\newcommand*{\figref}[1]{Fig.~\ref{#1}}
\newcommand*{\Figref}[1]{Figure~\ref{#1}}

\begin{document}

\title{Quantum Nonlinear Optics in Atomically Thin Materials}

\author{Dominik S. Wild}
\affiliation{Department of Physics, Harvard University, Cambridge, MA 02138, USA}

\author{Ephraim Shahmoon}
\affiliation{Department of Physics, Harvard University, Cambridge, MA 02138, USA}

\author{Susanne F. Yelin}
\affiliation{Department of Physics, Harvard University, Cambridge, MA 02138, USA}
\affiliation{Department of Physics, University of Connecticut, Storrs, CT 06269, USA}

\author{Mikhail D. Lukin}
\affiliation{Department of Physics, Harvard University, Cambridge, MA 02138, USA}

\begin{abstract}
 We show that a nonlinear optical response associated with a resonant, atomically thin material can be dramatically enhanced by placing it in front of a partially reflecting mirror, rendering otherwise weakly nonlinear systems suitable for experiments and applications involving quantum nonlinear optics. Our approach exploits the nonlinear response of long-lived polariton resonances that arise at particular distances between the material and the mirror. The scheme is entirely based on free-space optics, eliminating the need for cavities or complex nanophotonic structures. We analyze a specific implementation based on exciton-polariton resonances in two-dimensional semiconductors and discuss the role of imperfections and loss.
\end{abstract}

\maketitle

The realization of strong nonlinear interactions between individual light quanta (photons) has been a long-standing goal in optical science and engineering that is both of fundamental and technological significance~\cite{Chang2014}. While in conventional optical materials the nonlinearity at light powers corresponding to single photons is negligibly weak, remarkable advances  have been recently made towards realizing this goal. One promising approach to quantum nonlinear optics is based on quantum emitters confined to cavities or nanophotonic structures that greatly enhance light--matter interactions. Proof-of-principle experiments have been carried out with neutral atoms~\cite{Vetsch2010,Alton2011,Thompson2013}, quantum dots~\cite{Lodahl2015}, quantum wells~\cite{Munoz-Matutano2017,Delteil2018}, and color centers in diamond~\cite{Gao2015,Sipahigil2016}. At the same time, experiments with cold gases~\cite{Hammerer2010}, ensembles of solid state quantum emitters~\cite{DeRiedmatten2008}, and excitons in transition metal dichalcogenides (TMDs)~\cite{Back2018,Scuri2018} have demonstrated strong light--matter coupling without the need for nanophotonic structures. This is achieved via spatially delocalized optical excitation, which, however, reduces the nonlinearity, thereby rendering the system effectively linear at the level of individual photons. A number of solutions to this challenge have been proposed, for example, exploiting Rydberg blockade to induce strong, nonlocal interactions between ultracold atoms that result in strong photon--photon interactions~\cite{Saffman2010}. This approach has been applied to realize photon blockade~\cite{Pritchard2010,Peyronel2012,Baur2014}, two- and three-photon bound states~\cite{Liang2018} and symmetry protected collisions between strongly interacting photons~\cite{Thompson2017}. Extending such techniques to the domain of integrated solid-state systems is an outstanding challenge. 

\begin{figure}[t]
  \centering
  \includegraphics[width=\columnwidth]{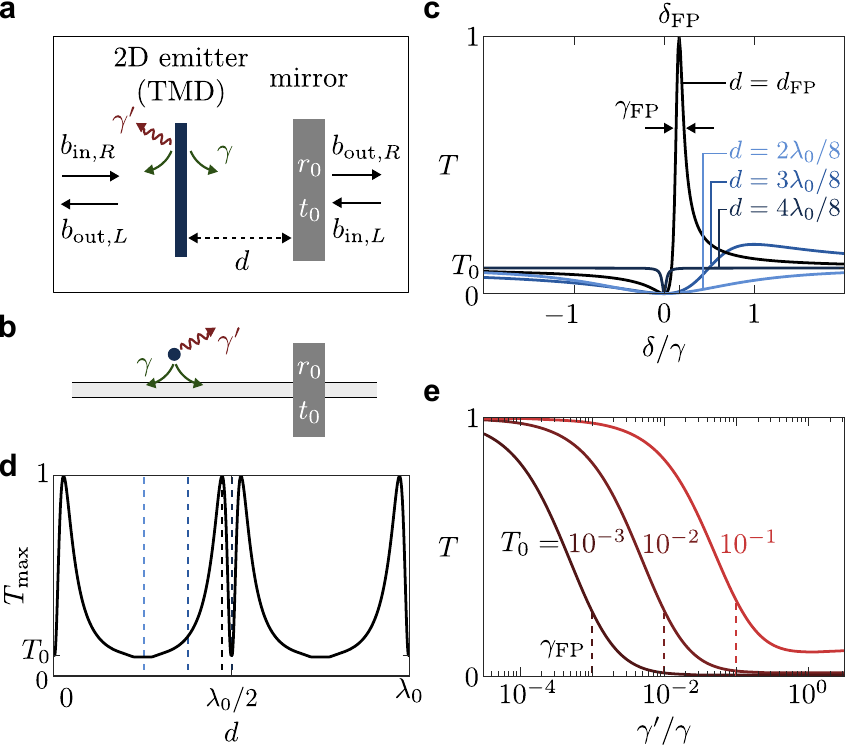}
  \caption{(a)~An atomically thin (2D) emitter is positioned at a distance $d$ in front of a partially reflecting mirror with reflection and transmission coefficients $r_0$ and $t_0$, respectively. The free-space radiative decay rate is given by $\gamma$, while $\gamma'$ denotes the loss rate. The setup is formally equivalent to (b), where the two-dimensional emitter is replaced by an atom and a one-dimensional waveguide takes the role of the free-space plane-wave mode. (c)~Transmission spectrum in the limit $\gamma \ll \nu_\mathrm{FSR}$ for $\gamma' = 0$ and $T_0 = 0.1$ at various distances $d$. (d)~Maximum transmission as a function of $d$. The dashed lines indicate the distances for which the full spectra are shown in (c). The maximum transmission is periodic in $d$ with period $\lambda_0/2$. (e)~Transmission at the Fabry--P\'erot resonance ($d = d_\mathrm{FP}$, $\delta = \delta_\mathrm{FP}$) as a function of the loss rate. Transmission is high provided $\gamma' \ll \gamma_\mathrm{FP} = T_0 \gamma$.}
  \label{fig:fig1}
\end{figure}

This Letter describes a novel approach to quantum nonlinear optics, which makes use of resonant, atomically thin materials. One example of such a material is a two-dimensional semiconductor such as a TMD monolayer, which supports tightly bound, optically active excitons~\cite{Wang2012}. In free space, excitons with zero in-plane momentum decay with a radiative rate $\gamma$, emitting a plane wave to either side of the TMD. Interactions between excitons render the system nonlinear, giving rise to a shift of the two-exciton state relative to the noninteracting case. However, in practice this nonlinearity is very weak, requiring a large number of excitons to create a resolvable shift. To enhance the nonlinear optical response, in our approach, the TMD is placed in front of a partially reflecting broadband mirror as shown in \figref{fig:fig1}a. When the separation between the TMD and the mirror is close to a half-integer multiple of the exciton resonance wavelength, the light emitted by the TMD towards the left (\figref{fig:fig1}a) and the light reflected by the mirror destructively interfere, which leads to significant suppression of the radiative linewidth, enhancement of the exciton lifetime and an associated enhancement of the optical nonlinearity. This effect can be understood in terms of the formation of long-lived exciton-polaritons, with a substantial excitonic component, which become very sensitive to nonlinear frequency shifts arising from exciton--exciton interactions.   

Before proceeding, we note an important connection with the single-atom system shown in \figref{fig:fig1}b, where a point-like emitter is coupled to a one-dimensional waveguide. Both systems can be thought of as a single-channel scattering problem with an emission rate $\gamma$ into the channel of interest and a loss rate $\gamma'$. In the case of a single atom, $\gamma'$ is typically dominated by emission into unconfined free-space modes. By contrast, conservation of in-plane momentum prevents scattering into undesired channels for a TMD, and loss only emerges due to nonradiative decay and material imperfections such as disorder. The relevant condition of low loss, $\gamma' \ll \gamma$, has already been demonstrated in high quality samples~\cite{Back2018,Scuri2018}. Similar considerations apply to other two-dimensional systems such as ordered arrays of trapped atoms with subwavelength spacing~\cite{Bettles2016,Shahmoon2017}. 

The key idea of this work can be understood by  first considering the linear response of a TMD in free space. The amplitude reflection coefficient close to an excitonic resonance is given by the complex Lorentzian $r_\mathrm{TMD}(\delta) = - i (\gamma/2)/[\delta + i (\gamma + \gamma')/2]$, where $\delta$ denotes the detuning from resonance~\cite{Zeytinoglu2017}. Strikingly, the TMD acts as a perfect reflector at zero detuning in the absence of losses despite being much less than a wavelength thick.  The vanishing transmission is the result of resonant scattering into a single channel, where the incident field destructively interferes with the scattered field. The effect has been discussed in a variety of other contexts including a single atom coupled to a one-dimensional waveguide~\cite{Chang2007}, classical plasmonic resonators~\cite{GarciadeAbajo2007}, and ordered arrays of atoms~\cite{Bettles2016,Shahmoon2017}. 
 
A nearby mirror significantly modifies the optical response. The intensity transmission coefficient $T$ can be computed by summing over all multiple reflections between the TMD and the mirror~\cite{SupMat}. As shown in \figref{fig:fig1}c and d, the transmission spectrum strongly depends on the distance $d$. In particular, perfect transmission is only attainable at specific distances close to half-integer multiples of the exciton transition wavelength $\lambda_0$. This distinction from a conventional Fabry--P\'erot resonator originates from the frequency dependence of the TMD. Perfect transmission through a Fabry--P\'erot resonator occurs when two conditions are met: The round trip phase is an integer multiple of $2 \pi$ and the reflection coefficients of the two mirrors are equal. Applied to our system, the latter condition may be stated as $|r_\mathrm{TMD}(\delta)|^2 = |r_0|^2$, which sets the detuning at which the Fabry--P\'erot resonance occurs, $\delta_\mathrm{FP} = \pm (\gamma/2) \sqrt{T_0 /R_0}$. The former condition then determines the allowed distances $d_\mathrm{FP}$ according to the relation $r_0 e^{2 i k_\mathrm{FP} d_\mathrm{FP}} = - R_0 \pm i \sqrt{R_0 T_0}$, where $k_\mathrm{FP}$ is the wavenumber corresponding to the resonance frequency. Here, $r_0$ (assumed to be real and negative) and $t_0$ denote the  amplitude reflection and transmission coefficient of the mirror, while $R_0 = |r_0|^2$ and $T_0 = |t_0|^2$ refer to the respective intensity coefficients. We may estimate the width $\gamma_\mathrm{FP}$ of the high-transmission resonance by considering the phase accumulated by a photon during $N \approx 1/T_0$ round trips before it is transmitted through the mirror. If the photon is detuned by $\Delta$ from the resonance, it accumulates an additional propagation phase $\varphi_\mathrm{prop}(\Delta) = 2 N d \Delta / c$, where $c$ is the speed of light. Furthermore, the reflection phase imparted by the TMD is modified by $\varphi_\mathrm{TMD}(\Delta) \approx 2 N \Delta / \gamma$. The width of the resonance follows from $\varphi_\mathrm{prop}(\gamma_\mathrm{FP}) + \varphi_\mathrm{TMD}(\gamma_\mathrm{FP}) \approx 1$. In the limit $\varphi_\mathrm{prop} \gg \varphi_\mathrm{TMD}$, the phase from the TMD can be neglected and the system resembles a conventional Fabry--P\'erot resonator. We are interested in the opposite limit, $\varphi_\mathrm{prop} \ll \varphi_\mathrm{TMD}$, requiring that $\gamma$ be much smaller than the free spectral range $\nu_\mathrm{FSR} = c/(2d)$, which yields $\gamma_\mathrm{FP} \approx T_0 \gamma$. For a highly reflecting mirror, $\gamma_\mathrm{FP}$ is much smaller than the free-space linewidth $\gamma$. The narrow linewidth can be physically understood in terms of a long-lived polariton formed by an exciton and a photon localized between the TMD and the mirror. Spontaneous emission from the polariton is suppressed because the photonic component destructively interferes with the field emitted by the exciton~\cite{Hoi2015}. 

Since the polaritons are predominantly composed of excitonic degrees of freedom, the interaction between them is comparable to the interaction between excitons in the absence of a mirror. Yet, polaritons may interact over a much longer duration owing to their extended lifetime. If we denote the interaction energy between two excitons by $\chi$, we expect that a strong quantum nonlinearity can be observed if $\chi > T_0 \gamma$, corresponding to an effective enhancement of the nonlinearity by a factor $1/T_0$ compared to  free space. The quantum nonlinearity results in photon antibunching as the presence of a single polariton blocks transmission by shifting the Fabry--P\'erot resonance by more than its width. In what follows, we confirm this simplified analysis and show that this effect is robust to loss, provided the loss rate $\gamma'$ is smaller than $\gamma_\mathrm{FP}$, as is required to maintain near unity transmission (see \figref{fig:fig1}e).

\begin{figure}[t]
  \centering
  \includegraphics[width=\columnwidth]{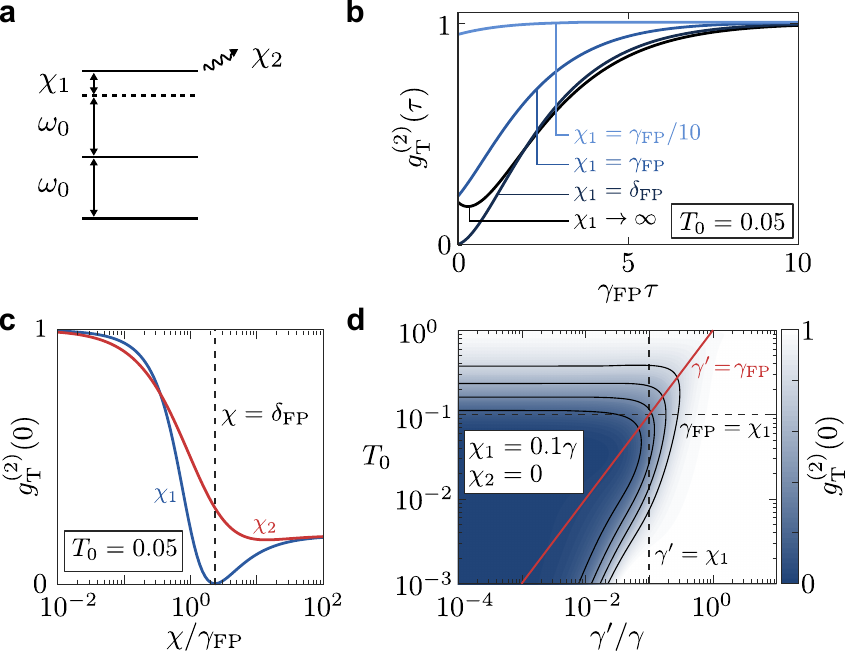}
  \caption{(a) Three lowest energy levels of the anharmonic oscillator used to model quantum nonlinear effects. Dispersive and dissipative nonlinearities are denoted by $\chi_1$ and $\chi_2$, respectively. (b) Second order correlation function $g_T^{(2)}(\tau)$ of the transmitted light at the Fabry--P\'erot resonance ($d = d_\mathrm{FP}$, $\delta = \delta_\mathrm{FP}$) for different values of $\chi_1$ while $\chi_2 = \gamma' = 0$. Pronounced photon antibunching is observed when $\chi_1 > \gamma_\text{FP}$. (c) Dependence of $g_T^{(2)}(0)$ on the strength of the nonlinearity. Only $\chi_1$ (blue) or $\chi_2$ (red) is varied, while the other parameter is set to zero.  (d) $g_T^{(2)}(0)$ as a function of $\gamma'$ and $T_0$. The experimentally relevant regime to observe strong antibunching is located below the horizontal dashed line ($\chi_1 > \gamma_\mathrm{FP}$) and above the red line ($\gamma_\mathrm{FP} > \gamma'$).}
  \label{fig:fig2}
\end{figure}

The above classical approach fully accounts for the linear response of the system but it is insufficient to capture quantum nonlinear effects. To this end, we quantize both the excitonic degrees of freedom and those of the electromagnetic field. The spatial mode of the excitons that couples to the light field is described by the bosonic annihilation and creation operators $a$ and $a^\dagger$. The internal dynamics of the excitons are governed by the Hamiltonian $H_0 = \omega_0 a^\dagger a + (\chi_1/2) a^\dagger a^\dagger a a$, where $\omega_0$ is the resonant frequency of the excitons and $\chi_1$ is the dispersive nonlinearity due to exciton---exciton interactions. A level diagram of the the three lowest energy states is shown in \figref{fig:fig2}a.  We employ an input--output formalism to eliminate the photonic degrees of freedom, upon which the equation of motion for a system operator $Q$ can be expressed in terms of the Heisenberg--Langevin equation~\cite{Gardiner1985,SupMat}
\begin{align}
  \label{eq:hl}
  \dot Q =  - &i \comm{Q}{H_0 + \frac{\gamma}{2} \im \left( r_0 e^{2 i k_0 d} \right) a^\dagger a + \Omega a^\dagger + \Omega^* a} \nonumber \\
  &+ \mathcal{D}[Q] + \mathcal{F}[Q],
\end{align}
where $k_0 = \omega_0/c$ and
\begin{equation}
  \Omega = \sqrt{\frac{\gamma}{2}} \left(  1 + r_0 e^{2 i k_0 d} \right) \langle b_\mathrm{in,R} \rangle + \sqrt{\frac{\gamma}{2}} t_0 e^{i k_0 d} \langle b_\mathrm{in,L} \rangle
\end{equation}
is the Rabi frequency. It is composed of a superposition of the input fields $b_\mathrm{in,R}$ and $b_\mathrm{in,L}$, illustrated in \figref{fig:fig1}a, which evolve as freely propagating photonic modes (as if the TMD and mirror were absent). The dissipative dynamics are described by
\begin{align}
  \mathcal{D}[Q] = &\left[ \gamma  + \gamma \re \left( r_0 e^{2 i k_0 d} \right) + \gamma' \right] \left( a^\dagger Q a - \frac{1}{2} \acomm{Q}{a^\dagger a} \right) \nonumber\\
  &+ \frac{\chi_2}{2} \left( a^\dagger a^\dagger Q a a - \frac{1}{2} \acomm{Q}{a^\dagger a^\dagger a a} \right).
\end{align}
In addition to the radiative decay rate $\gamma$ and the loss rate $\gamma'$, we include a nonlinear decay rate $\chi_2$. This rate accounts for the dissipative nonlinearity that may arise from nonradiative decay involving a pair of excitons, or from excitons scattering off each other into a spatial mode outside the mode of interest. Excitons may further be subject to pure dephasing, though the effect has been excluded here for the sake of clarity. We show in the supplemental material that a pure dephasing rate $\gamma_d$ affects the system in a qualitatively and quantitatively similar fashion to the loss rate $\gamma'$~\cite{SupMat}. Finally, the term $\mathcal{F}[Q]$ in \eqref{eq:hl} is a Langevin noise operator~\cite{SupMat}.

\Eqref{eq:hl} is valid under three assumptions: (\emph{i}) The Markov approximation applies, requiring that $\gamma \ll \omega_0$. This is typically justified for optical transitions and indeed holds for excitons in TMDs. (\emph{ii}) The photons initially occupy a coherent state that is uncorrelated with the excitons.  (\emph{iii}) Retardation can be neglected during a round trip of a photon traveling between the TMD and the mirror. We show in the supplemental material that this gives rise to the conditions $\gamma \ll  \nu_\text{FSR}$, $\gamma \ll \nu_\mathrm{FSR} \left[1 + \re\left( r_0 e^{2i k_0 d} \right) \right] / \im \left( r_0 e^{2i k_0 d} \right)$, and $\chi_{1,2} \ll \nu_\mathrm{FSR} \left[ 1 + \re \left( r_0 e^{2 i k_0 d} \right) \right]$, the first one being equivalent to the earlier condition that the Fabry--P\'erot resonance be dominated by the linewidth of the TMD. All three inequalities are easily met for $d$ on the order of an optical wavelength~\cite{SupMat}.

When these conditions are satisfied, the mirror affects the exciton dynamics in a rather simple way. It shifts the resonance frequency by $(\gamma/2) \im \left( r_0 e^{2 i k_0 d} \right)$ and modifies the radiative decay rate to $\tilde \gamma = \gamma \left[ 1 + \re \left( r_0 e^{2 i k_0 d} \right) \right]$. As alluded to previously, we may view the changes in energy and decay rate as a result of the formation of an exciton-polariton, where the excitonic degrees of freedom hybridize with a photonic mode that occupies the region between the TMD and mirror. At particular distances, the lifetime of the polariton can be significantly enhanced compared to an exciton in free space due to destructive interference between the radiation emitted by the exciton and the localized photonic mode.

After solving for the exciton dynamics governed by \eqref{eq:hl}, we can obtain the scattered field from the input--output relations~\cite{SupMat,Gardiner1985}
\begin{align}
  b_\text{out,L} &= t_0 e^{i k_0 d} b_\text{in,L} + r_0 e^{2 i k_0 d} b_\text{in,R}  + \! \sqrt{\frac{\gamma}{2}} \! \left( 1 \! + \! r_0 e^{2 i k_0 d} \right) a, \nonumber \\
  b_\text{out,R} &= t_0 e^{i k_0 d} b_\text{in,R} - \frac{t_0}{t_0^*} r_0^* b_\text{in,L} + \sqrt{\frac{\gamma}{2}} t_0 e^{i k_0 d} a.
  \label{eq:input_output}
\end{align}
These expressions have the simple interpretation that the output field arises from a superposition of the input fields with the field emitted by the TMD. Supposing that light is incident from the left, the reflection and transmission coefficients can be computed according to $R = \langle b_\text{out,L}^\dagger b_\text{out,L} \rangle / \langle  b_\text{in,R}^\dagger b_\text{in,R} \rangle$ and $T = \langle b_\text{out,R}^\dagger b_\text{out,R} \rangle / \langle  b_\text{in,R}^\dagger b_\text{in,R} \rangle$. For a weak input field, the coefficients computed in this manner agree with the classical result under the same conditions for which the Heisenberg--Langevin equation holds. The input--output relations also give us access to higher-order photon correlation functions such as the normalized two-time correlation function of the transmitted field,
\begin{equation}
  g_\mathrm{T}^{(2)}(\tau) = \frac{\langle b_\mathrm{out,R}^\dagger(0) b_\mathrm{out,R}^\dagger(\tau) b_\mathrm{out,R}(\tau) b_\mathrm{out,R}(0) \rangle}{\langle b_\mathrm{out,R}^\dagger(0) b_\mathrm{out,R}(0) \rangle \langle b_\mathrm{out,R}^\dagger(\tau) b_\mathrm{out,R}(\tau) \rangle}.
\end{equation}
Such correlation functions can be computed by expressing them in terms of two-time correlation functions of the excitonic operators, which can be related to one-time expectation values using the quantum regression theorem~\cite{Meystre2007}. Finally, we numerically evaluate the one-time expectation values from \eqref{eq:hl} in a truncated Fock space~\cite{SupMat}.

For the remainder of the discussion, we focus on the special case $d = d_\mathrm{FP}$, where the sharp Fabry--P\'erot resonance occurs. It is possible to neglect the difference between $k_\mathrm{FP}$ and $k_0$ under the same conditions that allowed us to ignore retardation. Hence, the resonance condition reads $r_0 e^{2 i k_0 d_\mathrm{FP}} = - R_0 \pm i \sqrt{R_0 T_0}$ and the radiative decay rate of the exciton-polariton is given by $\tilde \gamma = T_0 \gamma = \gamma_\mathrm{FP}$, consistent with the width of the transmission peak. To quantify the effect of line narrowing on the nonlinear dynamics, we plot $g_T^{(2)}(\tau)$ for different values of the dispersive nonlinearity $\chi_1$ in \figref{fig:fig2}b. Both the loss rate $\gamma'$ and the dissipative nonlinearity $\chi_2$ are taken to be zero, and we assume that a weak, monochromatic, coherent state resonant with the transmission peak ($\delta = \delta_\mathrm{FP}$) is incident on the TMD from the left. The figure clearly shows that $g_T^{(2)}(0)$ drops significantly below unity for $\chi_1 > \gamma_\mathrm{FP}$, confirming the presence a nonclassical state of light with strong photon antibunching~\cite{Meystre2007}. The effect may be understood by observing that the transmission peak of the single and two-exciton transitions are shifted relative to each other by $\chi_1$. If this shift exceeds the peak width $\gamma_\mathrm{FP}$, the second photon is reflected with high probability. The mechanism is closely related to polariton blockade in quantum well cavities, where the presence of a single polariton blocks subsequent photons from entering the cavity~\cite{Verger2006}. In the limit $\chi_1 \to \infty$, the transmission probability of the second photon is given by $T_0$, which explains the small but nonvanishing value of $g_T^{(2)}(0)$. By setting $\chi_1 = \delta_\mathrm{FP}$, it is possible to achieve $g_T^{(2)}(0) = 0$ because the transmission peak of the single exciton transition then perfectly matches a zero in transmission of the two-exciton transition (cf.~\figref{fig:fig1}c at zero detuning). We note that the line narrowing also affects the time scale over which antibunching is observed, the relevant time scale now being $1/\gamma_\mathrm{FP}$ rather than the free-space lifetime $1/\gamma$.

A dissipative nonlinearity can also give rise to photon antibunching. \Figref{fig:fig2}c shows $g_T^{(2)}(0)$ as a function of either $\chi_1$ or $\chi_2$, where the other parameter is set to zero. The two nonlinearities have a qualitatively similar effect on photon antibunching with the main difference being that the perfect antibunching dip at $\chi_1 = \delta_\mathrm{FP}$ is absent for the dissipative nonlinearity. While in both cases antibunching is caused by reduced transmission at the two-exciton transition, the dissipative nonlinearity accomplishes this by reducing the peak height rather than by shifting its position.

In order to observe antibunching in the presence of loss, it is necessary that the nonlinearity is large compared to not only $\gamma_\mathrm{FP}$ but also $\gamma'$. This is illustrated in \figref{fig:fig2}d, where we plot $g_T^{(2)}(0)$ as a function of both $\gamma'$ and $T_0$ for a fixed values of $\chi_1$ and $\chi_2$. In addition, we still require that $\gamma' < \gamma_\mathrm{FP}$ to ensure high transmission. These conditions may be summarized as
\begin{equation}
  \chi > T_0 \gamma > \gamma',
\end{equation}
where $\chi$ stands for $\chi_1$ or $\chi_2$.  In \figref{fig:fig2}d, these inequalities correspond to the region below the dashed horizontal line, to the left of the dashed vertical line, and above the diagonal red line. We point out that strong antibunching can be observed in the region $\gamma_\mathrm{FP} < \gamma' < \sqrt{T_0} \gamma$ below the red line. However, this region is of little practical relevance as it would be challenging to observe antibunching given the weak transmitted signal.

We next discuss the feasibility of our scheme with TMDs. Theoretical calculations~\cite{Shahnazaryan2017} have estimated the interaction energy between two excitons delocalized over an area $A$ in WS$_2$ (we expect it to be comparable in other TMDs) as $g \approx \SI{4}{\micro eV . \micro m^2} / A$~\footnote{We note that this value differs from existing estimates based on prior experimental measurements~\cite{Scuri2018}.  Our current understanding is that the discrepancy is due to dielectric screening by the substrate used in reference~\cite{Scuri2018}, which can be reduced, for instance, by suspending the samples. However, further work is required for a complete quantitative understanding of exciton--exciton interactions in TMDs.}. The parameter $\chi_1$ is obtained by computing the interaction energy of two excitons whose in-plane wavefunction is proportional to the electric field profile of the incident laser. For a Gaussian beam with waist $w_0$, we obtain $\chi_1 = g/(\pi w_0^2)$, where we assumed that the interaction is short ranged. The exciton transition in WS$_2$ occurs at $\lambda_0 \approx \SI{600}{nm}$~\cite{Chernikov2014}, which yields $\chi_1 \approx \SI{13}{\micro eV}$ for a diffraction limited spot ($w_0 = \lambda_0/2$). With the intrinsic linewidth given by $\gamma \approx \SI{3}{meV}$~\cite{Moody2015}, the transmission coefficient of the mirror must therefore satisfy $T_0 \lesssim 1/230$, which is experimentally feasible. In order to prevent the photonic mode from diverging over the course of $\sim\! 230$ roundtrips, the mirror must be curved with a radius of curvature that matches the incident beam. In practice, the enhancement will be limited by the loss rate $\gamma'$, which depends on the quality of the material. Values of $\gamma' < 0.1 \gamma$ have recently been achieved~\cite{Scuri2018}. While the current limiting factors are not fully understood, improved exciton properties have been observed in suspended devices~\cite{Scuri2018b}, and further advances in material quality will likely result in an additional decrease of $\gamma'$. We note that exciton dispersion places a lower bound on the loss rate. The bound is determined by the time scale $t_d = m w_0^2/(2 \hbar)$, $m$ being the exciton mass, after which the exciton wavefunction starts to significantly spread out, leading to a reduced overlap with the photonic mode, and thereby spoiling the Fabry--P\'erot resonance. For our purpose, the effect is expected to negligible since the associated rate $\hbar / t_d$ contributing to $\gamma'$ is found to be less than $\SI{1}{\micro eV}$.

The above considerations indicate that TMDs are a promising platform for exploring quantum nonlinear optical phenomena. In addition, our approach is compatible with methods that seek to increase the interaction strength between excitons in order to alleviate the requirements on the loss rate. It has been proposed that this may be accomplished using excited states of the exciton~\cite{Shahnazaryan2017} or by exploiting scattering resonances between excitons~\cite{Carusotto2010}. Interlayer excitons in TMD heterostructures offer an alternative route as their permanent dipole moment gives rise to much stronger interactions~\cite{Rivera2015}. Finally, applying a periodic potential to the TMD can enhance the interaction strength by increasing the local exciton density at the potential minima. The potential may be implemented with a modulation of the dielectric environment~\cite{Raja2017} or via the energy landscape arising from a moir\'e pattern in a heterostructure~\cite{Wu2018}.

In summary, we have demonstrated that a partially reflecting mirror can be used to dramatically enhance the lifetime of polaritons associated with a resonant two-dimensional material, which in turn enhances the sensitivity to weak nonlinearities. In contrast to other approaches of generating nonclassical states of light with weak nonlinearities, such as the unconventional photon blockade~\cite{Flayac2017,Ryou2018,Snijders2018,Vaneph2018}, our scheme does not require fine tuning between the nonlinearity and the loss rate. While we focused on TMDs, the approach is applicable to other emitters coupled to a single scattering channel such as two-dimensional arrays of trapped ultracold atoms~\cite{Bettles2016,Shahmoon2017}. In addition, the approach may be useful as a spectroscopic tool by narrowing emission lines, which does not require reaching the quantum nonlinear regime. The scheme can be naturally extended to multiple emitters such as several closely spaced layers of TMDs, which could replace the conventional mirror entirely~\cite{Chang2012}. Furthermore, future work could explore the crossover into the non-Markovian regime by moving the emitter sufficiently far away from the mirror such that retardation is no longer negligible~\cite{Dorner2002,Pichler2016}.

\begin{acknowledgments}
  We thank H.~Pichler, J.~Perczel, R.~E.~Evans, G.~Scuri, Y.~Zhou, P.~Kim, H.~Park, S.~Zeytino\u{g}lu, and A.~\.{I}mamo\u{g}lu for insightful discussions. This work was supported by NSF, CUA, AFOSR and the V. Bush Faculty Fellowship. 
\end{acknowledgments}

\bibliography{bibliography}
\end{document}